\documentclass[a4paper,10pt]{article}
\usepackage[utf8]{inputenc}
\usepackage{amsmath}
\usepackage{amsfonts}
\usepackage{color}
\usepackage{graphicx}
\usepackage{url}
\usepackage{bm}

\newcommand{\dd}{ {\rm d} }
\newcommand{\NN}{\mathcal{N}}
\newcommand{\UU}{\mathcal{U}} 

\newcommand{\g}{\mathcal{N}}

\newcommand{\mypartial}[1]{\frac{\partial}{\partial #1 }}

\newcommand{\x}{x} 
 \newcommand{\F}{\mathbf{F}}
\newcommand{\z}{z} 
\newcommand{\y}{y} \newcommand{\Y}{\mathbf{y}}
\newcommand{\messx}{\psi} 
\newcommand{\messd}{\phi} 
\newcommand{\mz}{\hat{Z}} \newcommand{\mZ}{\mathbf{\hat{Z}}}
\newcommand{\vz}{\bar{Z}} \newcommand{\vZ}{\mathbf{\bar{Z}}}
\newcommand{\mzp}{\hat{z}} \newcommand{\mZp}{\mathbf{\hat{z}}}
\newcommand{\vzp}{\bar{z}} \newcommand{\vZp}{\mathbf{\bar{z}}}
\newcommand{\mxa}{\hat{p}} \newcommand{\vxa}{\bar{p}}
\newcommand{\mx}{\hat{X}} \newcommand{\vx}{\bar{X}}
\newcommand{\mxp}{\hat{x}} \newcommand{\vxp}{\bar{x}}
\newcommand{\mX}{\mathbf{\hat{X}}} \newcommand{\vX}{\mathbf{\bar{X}}}
\newcommand{\mXp}{\mathbf{\hat{x}}} \newcommand{\vXp}{\mathbf{\bar{x}}}
\newcommand{\md}{\hat{D}} \newcommand{\vd}{\bar{D}}
\newcommand{\mdp}{\hat{d}} \newcommand{\vdp}{\bar{d}}

\newcommand{\p}{p}
\newcommand{\fm}{\hat{f}}
\newcommand{\fv}{\bar{f}}
\newcommand{\gm}{\hat{g}}
\newcommand{\gv}{\bar{g}}
\newcommand{\muilone}{\mu l \to i l}
\newcommand{\muiltwo}{\mu l , i}
\newcommand{\gilone}{\gamma l \to i l}

\usepackage{algorithm}
\usepackage{algorithmic}

\begin{document}

\title{Blind Sensor Calibration using Approximate Message Passing}
\author{Christophe~Schülke\thanks{C. Schülke is with Univ. Paris 7, Sorbonne Paris Cité, 75013 Paris, France, (e-mail: christophe.schulke@espci.fr).This work was supported by the ERC under the European Union's 7th Framework Programme Grant Agreement 307087-SPARCS and 
by Université franco-italienne.},
        Francesco~Caltagirone\thanks{F. Caltagirone and L. Zdeborov\'a are with Institut de Physique Théorique at CEA Saclay and CNRS URA 2306
91191 Gif-sur-Yvette, France, (e-mail: f.calta@gmail.com and lenka.zdeborova@gmail.com).}
        and~ Lenka~Zdeborov\'a\footnotemark[2]
%
}

%
%

\markboth{IEEE Transactions on Signal Processing}%
{Schülke \MakeLowercase{\textit{et al.}}: Blind Sensor Calibration using Approximate Message Passing }

\maketitle

\begin{abstract}
The ubiquity of approximately sparse data has led a variety of communities  to great interest in compressed sensing algorithms.
   Although these are very successful and well understood for linear measurements with additive noise, 
   applying them on real data can be problematic if imperfect sensing devices introduce deviations from this ideal 
   signal acquisition process, caused by sensor decalibration or failure.   
   We propose a message passing algorithm called calibration
   approximate message passing (Cal-AMP) that can treat a variety of such sensor-induced imperfections.
   In addition to deriving the general form of the algorithm, we numerically investigate two particular settings. 
   In the first, a fraction of the sensors is faulty, giving readings unrelated to the signal. 
   In the second, sensors are decalibrated and each one introduces a different multiplicative gain to the measurements.
   Cal-AMP shares the scalability of approximate message passing, allowing to treat big sized instances of these problems, and experimentally exhibits 
   a phase transition between domains of success and failure.
\end{abstract}



\section{Introduction}
\begin{figure}[h!]
  \centering
  \includegraphics[width=0.7\columnwidth]{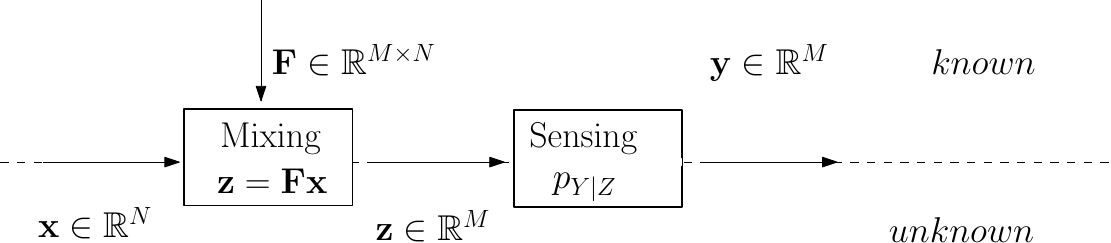}
  \caption{The generalized compressed sensing setting in GAMP \cite{Rangan10b}: the mixing step is followed by a sensing step, characterized 
  by the probability distribution $\p_{Y|Z}$.}
  \label{fig:mixAndSense}
  \end{figure}

  Compressed sensing (CS) has made it possible to algorithmically invert an underdetermined linear system, provided that the signal to recover is 
  sparse enough and that the mixing matrix has certain properties \cite{CandesRombergTao06}. 
  In addition to the theoretical interest raised by this discovery, CS is already used both in experimental research and 
  in real world applications, in which it can lead to significant improvements. 
  CS is particularly attractive for technologies in which an increase of the number of measurements is either impossible, as sometimes in 
    medical imaging \cite{lustig2007sparse, otazo2010combination}, or expensive, as in imaging devices that operate in certain wavelength \cite{duarte2008single}. 
  CS was extended to the setting in which the mixing process is followed by a sensing process which can be nonlinear 
  or probabilistic, as shown in Fig. \ref{fig:mixAndSense}, with an
  algorithm called the generalized approximate message passing (GAMP) \cite{Rangan10b}. 
  This has opened new applications of CS, such as phase retrieval \cite{schniter2012compressive}.
  
   \begin{figure}[h!]
   \centering
   \includegraphics[width=0.7\columnwidth]{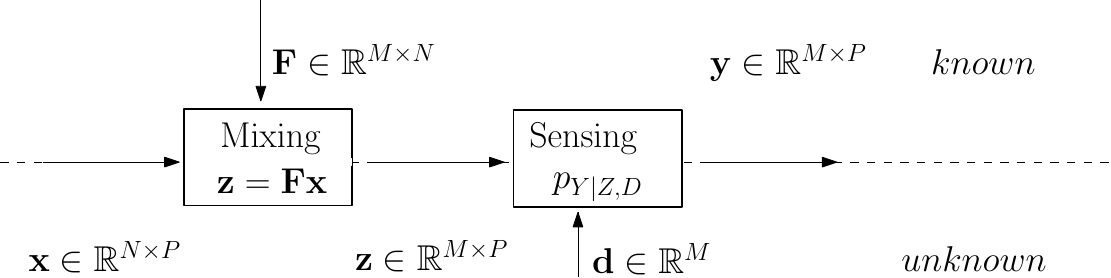}
   \caption{The
     blind calibration problem: the sensing process is known up to
     calibration parameters $\mathbf{d}$ that need to be recovered 
   jointly with the signal. For this to be possible, one generally needs to measure $P>1$ independent signals. 
   Note that the elements of $\mathbf{d}$ are characteristic of the sensing system and therefore do not depend on the signal measured.}
   \label{fig:mixAndSenseCalP}
   \end{figure}

  One issue that can arise in CS is a lack of knowledge or an uncertainty on the exact measurement process. 
  A known example is dictionary learning, where the measurement matrix $\mathbf{F}$ is not known. 
  The dictionary learning problem can also be solved with an AMP-based
  algorithm if the number $P$ of available signal samples grows as $N$
  \cite{krzakala2013phase}.
  
  A different kind of uncertainty is when the linear transformation $\mathbf{F}$,
  corresponding to the \textit{mixing} process, is known, 
  but the \textit{sensing} process is only known up to a set of parameters. 
  In some cases, it might be possible to estimate these parameters prior to the measurements in a \textit{supervised} sensor calibration process,
  during which one measures the outputs produced by \textit{known} training signals, and in this way estimate the parameters for each of the sensors.
  In other cases, this might not be possible or practical, and the parameters have to be estimated jointly with a set of unknown 
  signals:   this is known as the \textit{blind} sensor calibration problem. It is schematically shown on 
  Fig. \ref{fig:mixAndSenseCalP}.
  
  Some examples in which supervised calibration is impossible are given here:
  \begin{itemize}
   \item For supervised calibration to be possible, one must be able to measure a known signal.
   This might not always be the case: in radio astronomy for example, calibration is necessary~\cite{rau2009advances}, but the only possible 
   observation is the sky, which is only partially known.
   \item Supervised calibration is only possible when the system making the measurements is at hand, which might not 
   always be the case. Blind image deconvolution is an example of blind calibration in which the calibration parameters are 
   the coefficients of the imaging device's point spread function. It can easily be measured, but if we only have the blurred 
   images and not the camera, there is no other option than estimating the point spread function from the images themselves, 
   thus performing blind calibration~\cite{levin2009understanding}.
   \item For measurement systems integrated in embedded systems or smartphones, requiring a supervised calibration step 
   before taking a measurement might be possible, but is not user-friendly because it requires a specific calibration 
   procedure, which blind calibration does not. On the other hand, regular calibration might be necessary, as slow decalibration 
   can occur because of aging or external parameters such as temperature or humidity.
  \end{itemize}

  Several algorithms have been proposed for blind sensor calibration
  in the case of unknown multiplicative gains, relying on convex optimization \cite{GribonvalChardon11}
  or conjugate gradient algorithms \cite{shen2013conjugate}. 
  The Cal-AMP algorithm that we propose, and whose preliminary study
  was presented in~\cite{schulke2013blind}, is based on GAMP and is therefore not restricted to a specific output function. Furthermore, it has the same
  advantages in speed and scalability as the approximate message passing (AMP), and thus allows to treat
  problems with big signal sizes.

 \section{Blind sensor calibration: Model and notations}
  \subsection{Notations}
  In the following, vectors and matrices will be written using bold font. 
  The $i$-th component of the vector $\mathbf{a}$ will be written as $a_i$. 
  In a few cases, notations of the type $\mathbf{a}_i$ are used, in which case $\mathbf{a}_i$ is a vector itself,
  not the $i$-th component of vector $\mathbf{a}$.
  The complex conjugate of a complex number $x \in \mathbb{C}$ will be noted $x^*$, and its modulo $|x|$.
  The transpose (resp. complex transpose) of a real (resp. complex) vector $\mathbf{x}$ will be noted $\mathbf{x}^T$.
  The component-wise product between two vectors or matrices $\mathbf{a}$ and $\mathbf{b}$ will be noted $\mathbf{a} \odot \mathbf{b}$.
  The notations $\mathbf{a}^{-1}$, and $\frac{\mathbf{b}}{\mathbf{a}}$ are component-wise divisions, and $\mathbf{a}^2 = \mathbf{a}\odot \mathbf{a}$. 
  We will call a probability distribution function (pdf) on a matrix or vector variable $\mathbf{a}$ separable 
  if its components are independently distributed: $\p(\mathbf{a}) = \prod_i \p(a_i)$.
  Finally, we will write $\p(x) \propto f(x)$ if $\p$ and $f$ are proportional and we will write $x \sim \p_X(x)$ if 
  $x$ is a random variable with probability distribution function $\p_X$.
  
  \subsection{Measurement process}
  Let $\mathbf{\x}$ be a set of $P$ signals $\{ \mathbf{\x}_l , l=1\cdots P \}$ to be recovered
  and $N$ be their dimension: $\mathbf{\x}_l \in \mathbb{R}^N$.
  Each of those signals is sparse, meaning that only a fraction $\rho$ of their components is non-zero.
  
  The measurement process leading to $\mathbf{\y} \in \mathbb{R}^{M \times P}$ is shown in Fig. \ref{fig:mixAndSenseCalP}.
  In the first, linear step, the signal is multiplied by a matrix $\mathbf{F} \in \mathbb{R}^{M \times N}$ 
  and gives a variable $\mathbf{\z} \in \mathbb{R}^{M \times P}$
  \begin{align}
  \mathbf{\z} &= \mathbf{F} \mathbf{\x} ,
  \end{align}
  or, written component-wise
  \begin{align}
  \z_{\mu l} &= \sum_{i=1}^{N} F_{\mu i} \x_{il} \quad \quad {\rm for} \quad \mu=1\cdots M , \quad l=1\cdots P.
  \end{align} 
  We will refer to $\alpha = {M}/{N}$ as the measurement rate.
  In standard CS, the measurement~$\mathbf{\y}$ is a noisy version of
  $\mathbf{\z}$, and the goal is to reconstruct $\mathbf{\x}$ in the
  regime where the rate
  $\alpha <1$. 
  In the broader GAMP formalism, $\mathbf{\z}$ is only an intermediary
  variable that cannot directly be observed. The observation $\mathbf{\y}$ is a function of $\mathbf{\z}$, which is probabilistic 
  in the most general setting.
  
  In blind calibration, we add the fact that this function depends on an unknown parameter vector $\mathbf{d} \in \mathbb{R}^M$,
  such that the output function of each sensor is different, 
  \begin{equation}
    \Y \sim \p_{\mathbf{Y}| \mathbf{Z}, \mathbf{D}}(\Y|\mathbf{z},\mathbf{d}),
  \end{equation}
  with
  \begin{align}
  \p_{\mathbf{Y}| \mathbf{Z} , \mathbf{D}} (\mathbf{\y} | \mathbf{\z}, \mathbf{d}) &= \prod_{\mu = 1}^{M} \prod_{l=1}^{P} \p_{Y|Z,D}(\y_{\mu l}|\z_{\mu l},d_{\mu}) ,
  \end{align}
  and the goal is to jointly reconstruct $\mathbf{\x}$ and $\mathbf{d}$.

  
  
%

 \subsection{Properties AMP for compressed sensing}
It is useful to remind basic results known about the AMP algorithm for
compressed sensing \cite{DonohoMaleki09}. The AMP is derived on the
basis of belief propagation \cite{YedidiaFreeman03}. As is well known,
belief propagation on a loopy factor graph is not in general
guaranteed to give sensible results. However, in the setting of this paper, i.e. random iid
matrix $\mathbf{F}$ and signal with random iid elements of known
probability distribution, the AMP algorithm was proven to work in
compressed sensing in the the limit of large system size $N$ as long as
the measurement rate $\alpha \ge \alpha_{\rm CS}(\rho)$
\cite{DonohoMaleki09,BayatiMontanari10,DonohoMaleki10,KrzakalaMezard12}. The threshold
$\alpha_{\rm CS}(\rho)$ is a \textit{phase transition}, meaning that in
the limit of large system size, AMP fails with high probability
  up to the threshold $\alpha_{\rm CS}(\rho) \in ( \rho ,1)$ and
  succeeds with high probability above that threshold. 

  \subsection{Technical conditions}
  The technical conditions necessary for the derivation of the Cal-AMP
  algorithm and its good behavior are the following:
  \begin{itemize}
  \item Ideally, the prior distributions of both the signal, $\p_{\mathbf{X}}$, and the calibration parameters, $\p_{\mathbf{D}}$,
  are known, such that we can perform Bayes-optimal inference.
  As in CS, a mismatch between the real distribution and the assumed prior will in general 
  affect the performance of the algorithm. However, parameters of the real distribution can be learned with expectation-maximization
   and  improve performance~\cite{KrzakalaMezard12}.
  \item The Cal-AMP can be tested for an arbitrary operator~$\mathbf{F}$. However, in its derivation we assume that
    $\mathbf{F}$ is an iid random matrix, and that its elements are of order 
  $O(\frac{1}{\sqrt{N}})$, such that $\mathbf{\z}$ is $O(1)$ (given that $\mathbf{x}$ is $O(1)$). The
  mean of elements of $\mathbf{F}$ should be close to zero for the
    AMP-algorithms to be stable, in
    the opposite case the implementation has to be adjusted by some of
    the methods known to fix this issue \cite{caltagirone2014convergence}.
  \item The output function $\p_{\mathbf{Y}| \mathbf{Z} , \mathbf{D}}$ has to be separable, as well as the priors 
  on $\mathbf{\x}$ and $\mathbf{d}$. This condition could be relaxed by using techniques similar to those allowing to
  treat the case of structured sparsity in \cite{rangan2012hybrid}.
  \end{itemize}

  Under the above conditions we conjecture that in the limit of large
  system sizes the Cal-AMP algorithm matches the performance of the Bayes-optimal
  algorithm (except in a region of parameters where the Bayes-optimal fixed
  point of the Cal-AMP is not reached from an non-informed
  initialization, the same situation was described in compressed
  sensing \cite{KrzakalaMezard12}). This conjecture is based on
  the insight from the theory of spin glasses
  \cite{MezardMontanari09}, and it makes the Cal-AMP algorithm stand
  out among other possible extensions of GAMP that would take into
  account estimation of the distortion parameters. Proof of this
  conjecture is a non-trivial challenge for future work.

  \subsection{Relation to GAMP and some of its existing extensions}
  Cal-AMP algorithm can be seen as an extension of GAMP~\cite{Rangan10b}.

  Cal-AMP reduces to GAMP for the particular case
  of a single signal sample $P=1$.  Indeed, if the measurement $y_{\mu}$ depends 
  on a parameter $d_{\mu}$ via a probability distribution function $\p_{Y|Z,D}$, then $\p_{Y|Z}$ can
  be expressed by:
  \begin{equation}
   \p_{Y|Z}(y_{\mu}|z_{\mu}) = \int \dd d_{\mu} \p_D(d_{\mu})
   \p_{Y|Z,D}(y_{\mu}| z_{\mu},d_{\mu})\, .
  \end{equation}

   When, however, the number of signal samples is greater than one, $P>1$, the two algorithms differ:
   while GAMP treats the $P$ signals independently, leading to the same reconstruction performances 
   no matter the value of $P$, Cal-AMP treats them jointly. As our numerical results show,
   this can lead to great improvements in reconstruction performances, and can allow exact signal 
   reconstruction in conditions under which GAMP fails.

   One work on blind calibration that used a GAMP-based algorithm
   is~\cite{kamilov2013autocalibrated}, where the authors combine GAMP with expectation
maximization-like learning. That paper, however, considers a setting different
from ours in the sense that the unknown gains are on the signal
components not on the measurement components. Whereas both these cases
are relevant in practice, from an algorithmic point of view they are 
different.  

  Another work where distortion-like parameters are included and
  estimated with a GAMP-based algorithm is
  \cite{schniter2011message,nassar2014factor}.  Authors of this work consider
  two types of distortion-like parameters. Parameters $S$
  that are sample-dependent and hence their estimation is more related
  to what is done in the matrix factorization problem rather than to the
  blind calibration considered here. And binary parameters $b$ that are
  estimated independently of the main loop that uses GAMP. The 
  problem considered in that work requires a setting and a factor graph more complex
  that the one we considered here and it is far from transparent what to
  conclude about performance for blind calibration from the results presented in \cite{schniter2011message,nassar2014factor}.

\section{The Cal-AMP algorithm}
  In this section, we give details of the derivation of the approximate message passing algorithm 
  for the calibration problem (Cal-AMP). It is closely related to the AMP algorithm for CS \cite{DonohoMaleki09} 
  and the derivation was made using the same strategy as in~\cite{KrzakalaMezard12}.
  First, we express the blind sensor calibration problem as an inference problem, using Bayes' rule and an \textit{a priori} 
  knowledge of the probability distribution functions of both the signal and the calibration parameters. 
  From this, we obtain an \textit{a posteriori} distribution, which is peaked around the unique solution with high probability. 
  We write belief propagation equations that lead to an iterative
  update procedure of signal estimates. We realize that in the limit
  of large system size the algorithm
  can be simplified by working only with the means and variances of the
  corresponding messages. 
  Finally, we reduce the computational complexity of the algorithm by noting that the messages are
  perturbed versions of the local beliefs, which become the only quantities that need updating.
  
  \subsection{Probabilistic approach and belief propagation}
  We choose a probabilistic approach to solve the blind calibration
  problem, which has been shown to be very successful in CS. 
  The starting point is Bayes' formula that allows us to estimate the signal $\mathbf{\x}$ and the calibration parameters $\mathbf{d}$ from the knowledge
  of the measurements $\mathbf{\y}$ and the measurement matrix $\mathbf{F}$, assuming that $\mathbf{\x}$ and $\mathbf{d}$ are statistically independent,
  \begin{align}
  \p(\mathbf{\x},\mathbf{d}|\mathbf{\y},\mathbf{F}) &= \frac{\p_{\mathbf{X}}(\mathbf{\x})\p_{\mathbf{D}}(\mathbf{d})\p(\mathbf{\y}|\mathbf{F},\mathbf{\x},\mathbf{d})}{\p(\mathbf{\y}|\mathbf{F})} .
  \end{align}
  Using separable priors on $\mathbf{\x}$ and $\mathbf{d}$ as well as separable output functions, this posterior distribution becomes
  \begin{align}
  \p(\mathbf{\x}, \mathbf{d} | \mathbf{\y}, \mathbf{F}) = \frac{1}{Z} &\prod_{i,l=1}^{N,P} \p_X({\x}_{il}) \prod_{\mu =1}^{M} \p_D({d}_{\mu}) \times \nonumber \\
							  &\prod_{l,\mu=1}^{P,M} \p_{Y|Z,D}( {\y}_{\mu l}| {\z}_{\mu l}, {d}_{\mu} ) ,  \label{posterior}
  \end{align}
  where $Z$ is the normalization constant. 
  Even in the factorized form of (\ref{posterior}), uniform sampling
  from this posterior distribution becomes intractable with growing~$N$.

\begin{figure}
  \centering
  \vspace{-2mm}
  \includegraphics[width=6cm]{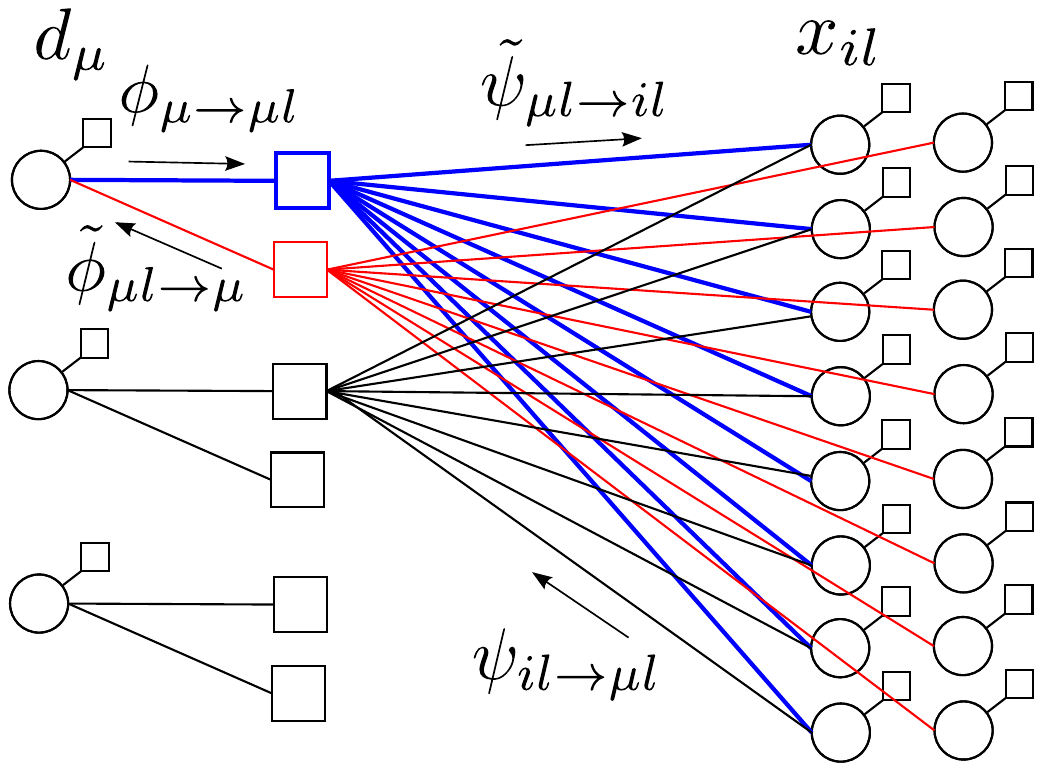}
  \caption{\label{factro_graph} Graphical model representing the posterior distribution (\ref{posterior}) of the blind calibration
    problem.  Here, the dimension of the signal is $N\!=\!8$, the
    number of sensors is $M\!=\!3$, and the number of signals used for
    calibration $P\!=\!2$. The variable nodes $\x_{il}$ and $d_\mu$ are
    depicted as circles, the factor nodes as squares (for clarity, only the three upper factor nodes are represented with all their links).}
  \label{factorGraph}
  \end{figure}

  Representing (\ref{posterior})  by the factor graph in Fig.
  \ref{factorGraph} allows us to use belief propagation for
  approximate sampling.
  As the factor graph is not a tree, there is no guarantee that
  running  belief propagation on it will lead to the correct
  results. Relying on the success of AMP in compressed sensing and
  the insight from the theory of spin glasses \cite{MezardMontanari09}, we conjecture belief
  propagation to be asymptotically exact in blind calibration as it is in CS.

  In belief propagation there are two types of pairs of messages: $(\messx,\tilde{\messx})$ and $(\messd,\tilde{\messd})$, 
  connected to the signal components and to the  calibration parameters respectively.
  Their updating scheme in the sum-product belief propagation is the
  following~\cite{YedidiaFreeman03}: for the $(\messd,\tilde{\messd})$ messages,
  \begin{align}
  \messd_{\mu \to \mu l}^t ({d}_{\mu}) &\propto \p_D({d_{\mu}}) \prod_{m \neq l} \tilde{\messd}_{\mu m \to \mu}^t ({d}_{\mu})  , \label{mess_n} \\
  \tilde{\messd}_{\mu l \to \mu}^{t+1} ({d}_{\mu}) &\propto \int \left( \prod_{i} {\rm d} {\x}_{il} \messx_{il \to \mu l}^t ({\x}_{il}) \right) \times \nonumber \\
      &\p_{Y|Z,D}({\y}_{\mu l} | \sum_i {F}_{\mu i} {\x}_{i l},{d}_{\mu} )  , \label{mess_n_tilde}
  \end{align}
  whereas for the  $(\messx,\tilde{\messx})$ messages,
  \begin{align}
  \messx_{il \to \mu l}^t ({\x}_{il}) &\propto \p_X({\x}_{il}) \prod_{\gamma \neq \mu} \tilde{\messx}_{\gamma l \to il}^t ({\x}_{il}) , \label{mess_m}\\
   \tilde{\messx}_{\mu l \to i l}^{t+1} ({\x}_{il})&\propto \int {\rm d} {d}_{\mu} \messd_{\mu \to \mu l}^t ({d}_{\mu}) \int \left( \prod_{j \neq i} {\rm d} {\x}_{jl} \messx_{j l \to \mu l}^t ({\x}_{jl}) \right) \times \nonumber \\
      &\p_{Y|Z,D}({\y}_{\mu l} | \sum_i {F}_{\mu i} {\x}_{i l},{d}_{\mu}) . \label{mess_m_tilde}
  \end{align}
  When belief propagation is successful, these messages converge to a fixed point,
  from which we obtain the marginal distribution of $\mathbf{x}$ sampled with (\ref{posterior}):
  \begin{align}
   \messx_{il}^t &\propto \p_X(x_{il}) \prod_{\gamma} \tilde{\messx}_{\gamma l \to il}^t(x_{il}). \label{eq:beliefsx}
  \end{align}
  These distributions are called beliefs, and from them we obtain the minimal mean square error  (MMSE) estimator:
  \begin{align}
   \hat{x}_{il}^{\rm MMSE} = \int {\rm d}x_{il} \, x_{il} \messx_{il}(x_{il}). \label{eq:MMSE}
  \end{align}

  \subsection{Simplifications in the large $N$ limit}
  The above update equations are still intractable, given the fact that in general, $\x_{il}$ and $d_\mu$ are continuous variables. 
  In the large $N$ limit, the problem can be greatly simplified by making leading-order expansions of certain quantities as a function
  of the matrix elements $F_{\mu i}$, that are of order $1/\sqrt{N}$. The notation $O(F_{\mu i})$ is therefore 
  equivalent to $O(1/\sqrt{N})$.
  
  This allows to pass messages that are estimators of variables and of their uncertainty, instead of full probability distributions.
  The table in Fig.~{\ref{table:notations} is a summary of notations used and their significations: estimators of variables are noted with a hat, whereas 
  their uncertainties are noted with a bar.
  \begin{figure}
  \centering
  \begin{tabular}{|c|c|c|c|c|c|c|}
   \hline
   variable &\multicolumn{2}{|c|}{$x$} & \multicolumn{2}{|c|}{$z$} & \multicolumn{2}{|c|}{$d$} \\
   \hline
   mean & $\mx$ & $\mxp$ & $\mz$ & $\mzp$ & $\md$ & $\mdp$ \\
   variance & $\vx$ & $\vxp$ & $\vz$ & $\vzp$ & $\vd$ & $\vdp$ \\
   \hline
  \end{tabular}
  \caption{Notations of the estimators and uncertainty estimators of the variables to be inferred. 
  Upper case letters represent estimations obtained from the most recent estimates of the other variables, 
  lower case letters are estimates taking into account the prior (for $\x$ and $d$) and the data (for $z$).
  }
  \label{table:notations}
  \end{figure}
 
  The messages can then be expressed in simpler ways by using Gaussians.
  As these will be ubiquitous in the rest of the paper, let us introduce the notation
  \begin{align}
  \g({x};{R},{\Sigma}) &= \frac{e^{-\frac{ (x-R)^2}{2 \Sigma}}}{\sqrt{2 \pi \Sigma}} ,
  \end{align}
  and note the expression of the following derivative:
  \begin{align}
   \mypartial{R} \g(x;R,\Sigma) &= \frac{x-R}{\Sigma} \g(x;R,\Sigma). \label{eq:gaussianderivate}
  \end{align}

  We will also use convolutions of a function $g$, with optional parameters $\{ u \}$, with a Gaussian
  \begin{align}
  f_k^g(R,\Sigma,\{ u \}) &= \int {\rm d}x \, x^k g(x,\{ u \})  \NN(x;R,\Sigma) ,  \nonumber \\ 
  \fm^g(R,\Sigma,\{ u \}) &= \frac{f_1^g(R,\Sigma,\{ u \})}{f_0^g(R,\Sigma,\{ u \})} , \label{gaussConvs} \\
  \fv^g(R,\Sigma,\{ u \}) &= \frac{f_2^g(R,\Sigma,\{ u \})}{f_0^g(R,\Sigma,\{ u \})} - | \fm^g(R,\Sigma,\{ u \}) |^2 , \nonumber
  \end{align}
  and from~(\ref{eq:gaussianderivate}), we obtain the relations
  \begin{align}
  \mypartial{R} f_k^g(R,\Sigma,\{ u \}) &= \frac{ f_{k+1}^g(R, \Sigma,\{ u \}) - R f_k^g(R,\Sigma,\{ u \})}{\Sigma} ,  \label{eq:fderivatives} \\
    \Sigma \, \mypartial{R} \fm^g({R},{\Sigma},\{ u \}) &= \fv^g({R},{\Sigma},\{ u \}) . \nonumber
  \end{align}

Let us show how simplifications come about in the large $N$ limit. Both in (\ref{mess_n_tilde}) and in (\ref{mess_m_tilde}), 
the term $p_{Y|Z,D}(y_{\mu l}|\z_{\mu l} = \sum_i F_{\mu i} \x_{il}, d_{\mu})$ appears. $\z_{\mu l}$ is a sum of the $N$ random 
variables $F_{\mu i} \x_{il}$, and each $\x_{il}$ is distributed according to the distribution $\messx_{il \to \mu l}^t(\x_{il})$.
Let us  call $\mxp_{il \to \mu l}$ and $\vxp_{il \to \mu l}$ the means and variances of these distributions,
  \begin{align}
  {\mxp}_{il \to \mu l}^t &= \int {\rm d}\x_{il} \, {\x}_{il} \messx_{il \rightarrow \mu l}^t ({\x_{il}}) , \\ 
  {\vxp}_{il \to \mu l}^t &= \int {\rm d}\x_{il} \, {\x}_{il}^2 \messx_{il \rightarrow \mu l}^t ({\x_{il}}) - ({\mxp}^t_{il \to \mu l})^2.
  \label{eq:meansNaive}
  \end{align}
In the $N \to \infty$ limit, we can use the central limit theorem, as the assumption of independence of the 
variables is already made when writing the belief propagation equations.
Then, $\z_{\mu l}$ has a normal distribution with means and variances:
  \begin{align}
  {\mz}_{\mu l}^{t+1} &= \sum_i {F}_{\mu i} {\mxp}_{il \to \mu l}^t , \\
  {\vz}_{\mu l}^{t+1} &= \sum_i {F}_{\mu i}^2 {\vxp}_{il \to \mu l}^t .
  \end{align}

  In (\ref{mess_n_tilde}) we therefore obtain
  \begin{align}
   \tilde{\messd}_{\mu l \to \mu}^{t+1} (d_{\mu})  &\propto \int {\rm d}z_{\mu l} \NN(\z_{\mu l}; \mz_{\mu l}^{t+1} , \vz_{\mu l}^{t+1})  \p_{Y|Z,D}(y_{\mu l}|\z_{\mu l},d_{\mu}) \nonumber \\
					     &\propto f_0^Z(\mz_{\mu l}^{t+1},\vz_{\mu l}^{t+1},\y_{\mu l},d_{\mu}) ,
  \label{simple_tilde_n}
  \end{align}
  where $f_0^Z(\mz,\vz,\y,d)$ is a lighter notation for $f_0^{\p_{Y|Z,D}}(\mz,\vz,\{y,d\})$ given by the formula in (\ref{gaussConvs}).
  
  For the $\messd$ messages, we obtain that
  \begin{equation}
  \messd_{\mu \to \mu l}^t(d_{\mu}) \propto \p_D(d_{\mu}) \prod_{m \neq l} f_0^Z(\mz_{\mu m}^{t},\vz_{\mu m}^{t},\y_{\mu m},d_{\mu}) .
  \label{simple_n}
  \end{equation}

  The same procedure can be applied to the $\tilde{\messx}$ messages, the only difference being that
  $\x_{il}$ is fixed, leading to
  \begin{align}
  \tilde{\messx}_{\mu l \to il}^{t+1}({\x}_{il}) 
   &\propto \int {\rm d}{d}_{\mu}   f_0^Z(\mz_{\muilone}^{t+1} + F_{\mu i} \x_{il}, \vz_{\muilone}^{t+1}, \y_{\mu l},d_{\mu}) \times  \nonumber   \\
   & \p_D({d}_{\mu}) \prod_{m \neq l} f_0^Z(\mz_{\mu m}^{t},\vz_{\mu m}^{t},\y_{\mu m},d_{\mu}) ,
  \label{eq:exactmesstildex}
  \end{align}
  with
  \begin{align}
  {\mz}_{\muilone}^{t+1} &= \sum_{j \neq i} {F}_{\mu j} {\mxp}_{jl \to \mu l}^t , \\
  {\vz}_{\muilone}^{t+1} &= \sum_{j \neq i} {F}_{\mu j}^2 {\vxp}_{jl \to \mu l}^t .
  \end{align}
 
  In analogy to the functions defined in (\ref{gaussConvs}), we introduce the functions 
  of the $P$-dimensional vectors $\mathbf{\mZ}$, $\mathbf{\vZ}$ and $\mathbf{\y}$:
  \begin{align}
  g_{k} (\mathbf{\mz}, \mathbf{\vz},\mathbf{\y}) &= \int \dd d_{\mu} \p_D(d_{\mu}) f_k^Z (\mz_{1},\vz_{1},y_{1},d_{\mu}) \times \nonumber \\
							&\prod_{m=2}^P f_0^Z (\mz_{m},\vz_{m},y_{m},d_{\mu}) ,  \nonumber \\
  \gm (\mathbf{\mz}, \mathbf{\vz},\mathbf{\y}) &= \frac{g_{1} (\mathbf{\mz}, \mathbf{\vz},\mathbf{\y})}{g_{0} (\mathbf{\mz}, \mathbf{\vz},\mathbf{\y})},  \label{eq:gfunctions}\\
  \gv (\mathbf{\mz}, \mathbf{\vz},\mathbf{\y}) &= \frac{g_{2} (\mathbf{\mz}, \mathbf{\vz},\mathbf{\y})}{g_{0} (\mathbf{\mz}, \mathbf{\vz},\mathbf{\y})} - | \gm (\mathbf{\mz}, \mathbf{\vz},\mathbf{\y}) |^2 , \nonumber
 \end{align}
 and as for the functions $f_k$, we can use (\ref{eq:gaussianderivate}) to show that
 \begin{align}
  \mypartial{\mz_1} g_{k}(\mathbf{\mz}, \mathbf{\vz},\mathbf{\y}) &= \frac{g_{k+1}(\mathbf{\mz}, \mathbf{\vz},\mathbf{\y})- \mz_1 g_{k}(\mathbf{\mz}, \mathbf{\vz},\mathbf{\y})}{\vz_1} , \label{eq:gderivatives} \\
  \mypartial{\mz_1} \gm(\mathbf{\mz}, \mathbf{\vz},\mathbf{\y}) &= \frac{\gv (\mathbf{\mz}, \mathbf{\vz},\mathbf{\y})}{\vz_1}. \nonumber
 \end{align}

With these functions, we define new estimators $\mzp$ and $\vzp$ of $z$:
\begin{align}
 \mzp_{\mu l \to il}^{t+1} &\equiv \gm(\mathbf{\mz}_{\muiltwo}^{t+1},\mathbf{\vz}_{\muiltwo}^{t+1}, \mathbf{\y}_{\mu l}) , \\
 \vzp_{\mu l \to il}^{t+1} &\equiv \gv(\mathbf{\mz}_{\muiltwo}^{t+1},\mathbf{\vz}_{\muiltwo}^{t+1}, \mathbf{\y}_{\mu l}).
\end{align}
Here, we use $\mathbf{\mz}_{\muiltwo}^{t+1}$ as a compact notation for the $P$-dimensional vector $\{ \mz_{\muilone}^{t+1} , \{ \mz_{\mu m}^{t} \}_{m \neq l} \}$, 
similarly for $\mathbf{\vz}_{\muiltwo}^{t+1}$, and $\mathbf{\y}_{\mu l}$ for the $P$-dimensional vector $\{\y_{\mu l}, \{ \y_{\mu m} \}_{m \neq l} \} \}$.
In appendix~\ref{app:A}, we show how we can obtain the following approximation for the $\messx$ messages:
  \begin{equation}
  \messx_{\mu l \to il}^t ({\x}_{il}) \propto \p_X({\x}_{il}) \left( \NN({\x}_{il};\mx_{il \to \mu l}^t,\vx_{il \to \mu l}^t) + O(\frac{x_{il}^3}{\sqrt{N}}) \right),
  \label{simple_m}
  \end{equation}
with
\begin{align}
  \vx_{il \to \mu l}^{t+1} &= \left( \sum_{\gamma \neq \mu} \frac{F_{\gamma i}^2 \left( \vz_{\gilone}^{t+1} - \vzp_{\gamma l \to i l}^{t+1} \right)}{(\vz_{\gilone}^{t+1})^2} \right)^{-1},\\
  \mx_{il \to \mu l}^{t+1} &= \vx_{il \to \mu l}^{t+1} \sum_{\gamma \neq \mu} \frac{F_{\gamma i}}{\vz_{\gilone}^{t+1}} \left( \mzp_{\gamma l \to i l}^{t+1} - \mz_{\gilone}^{t+1} \right). \nonumber
\end{align} 
In the $N\to \infty$ limit, the means and variances of $\messx_{\mu l \to il}(\x_{il})$ are therefore given by:
 \begin{align}
  {\mxp}_{il \to \mu l}^t &= \fm^X \left( \mx_{il \to \mu l}, \vx_{il \to \mu l} \right) , \\
  {\vxp}_{il \to \mu l}^t &= \fv^X \left( \mx_{il \to \mu l}, \vx_{il \to \mu l} \right),
  \label{eq:means}
  \end{align}
  where we have simplified the notations $\fm^{\p_X}$ and $\fv^{\p_X}$ to $\fm^{X}$ and $\fv^{X}$. 

%
%
  \subsection{Resulting update scheme}
  The message passing algorithm obtained by those simplifications is an iterative update scheme for
  means and variances of Gaussians. Given the variables at a time step $t$, the first step consists in producing estimates of $\mathbf{z}$:
  \begin{align}
  {\vz}_{\muilone}^{t+1} &= \sum_{j \neq i} {F}_{\mu j}^2 {\vxp}_{jl \rightarrow \mu l}^t, \quad {\vz}_{\mu l}^{t+1} = \sum_{j} {F}_{\mu j}^2 {\vxp}_{jl \rightarrow \mu l}^t, \\
  {\mz}_{\muilone}^{t+1} &= \sum_{j \neq i} {F}_{\mu j} {\mxp}_{jl \to \mu l}^t, \quad   {\mz}_{\mu l}^{t+1} = \sum_{j} {F}_{\mu j} {\mxp}_{jl \to \mu l}^t.
  \end{align}
  This step is purely linear and produces estimates $\mz_{\mu l}^{t+1}$ of $z_{\mu l}$ along with estimates of the incertitude $\vz_{\mu l}^{t+1}$. 
  The corresponding variables with arrows exclude one term of the sum, and are necessary in the belief propagation algorithm.
  
  The next step produces a new estimate of $\mathbf{z}$ from a nonlinear function of the previous estimates and the measurements $\mathbf{y}$:  
  \begin{align}
  \vzp_{\mu l \to i l}^{t+1} &= \gv \left( \mathbf{\mz}_{\muiltwo}^{t+1},\mathbf{\vz}_{\muiltwo}^{t+1}, \mathbf{\y}_{\mu l} \right) ,  \label{eq:mess_z_bar} \\
  \mzp_{\mu l \to i l}^{t+1} &=  \gm \left( \mathbf{\mz}_{\muiltwo}^{t+1},\mathbf{\vz}_{\muiltwo}^{t+1}, \mathbf{\y}_{\mu l} \right). \label{eq:mess_z_hat}
  \end{align}
  
  Next, the previous estimates of $\mathbf{z}$ are used in a linear step producing new estimates of $\mathbf{x}$:  
  \begin{align}
  \vx_{il \to \mu l}^{t+1} &= \left( \sum_{\gamma \neq \mu} \frac{F_{\gamma i}^2 \left( \vz_{\gilone}^{t+1} - \vzp_{\gamma l \to i l}^{t+1} \right)}{(\vz_{\gilone}^{t+1})^2} \right)^{-1} , \\
  \mx_{il \to \mu l}^{t+1} &= \vx_{il \to \mu l}^{t+1} \sum_{\gamma \neq \mu} \frac{F_{\gamma i}}{\vz_{\gilone}^{t+1}} \left( \mzp_{\gamma l \to i l}^{t+1} - \mz_{\gilone}^{t+1} \right).
  \end{align}
  
  Finally, a nonlinear function is applied to these estimates in order to take into account the sparsity constraint:
  \begin{align}
 {\mxp}_{il \rightarrow \mu l}^{t+1} &= \fm^X \left( \mx_{il \to \mu l}^{t+1}  , \vx_{il \to \mu l}^{t+1}  \right),  \label{mess_a}\\
  {\vxp}_{il \rightarrow \mu l}^{t+1} &= \fv^X \left( \mx_{il \to \mu l}^{t+1}  , \vx_{il \to \mu l}^{t+1} \right).  \label{mess_v}
  \end{align}


  \subsection{TAP algorithm with reduced complexity}
  In the previous message passing equations, we have to update $O(MPN)$ variables at each iteration. It turns out that this is not necessary, considering that
  the final quantities we are interested in are not the messages $\mxp_{il \to \mu l}$, but rather the local beliefs $\mxp_{i l}$. With that in mind,
  we can use again the fact that $F_{\mu i}$ is small to make expansions that will reduce the number of variables to actually update.
  Similarly to the messages (\ref{eq:mess_z_bar}), (\ref{eq:mess_z_hat}), (\ref{mess_a}) and (\ref{mess_v}), we define following quantities: 
  \begin{align}
  \mxp_{il}^t &= \fm^X \left( \mx_{il}^t,\vx_{il}^t \right), & \mzp_{\mu l}^t &= \gm \left( \mathbf{\mz}_{\mu l}^{t},\mathbf{\vz}_{\mu l}^{t}, \mathbf{\y}_{\mu l}  \right) , \nonumber \\
  \vxp_{il}^t &= \fv^X \left( \mx_{il}^t,\vx_{il}^t \right), &  \vzp_{\mu l}^t &= \gv \left( \mathbf{\mz}_{\mu l}^{t},\mathbf{\vz}_{\mu l}^{t}, \mathbf{\y}_{\mu l}   \right) , \label{eq:beliefs}
  \end{align}
  with
\begin{align}
  \vx_{il}^{t} &= \left( \sum_{\gamma } \frac{F_{\gamma i}^2 \left( \vz_{\gilone}^{t} - \vzp_{\gamma l \to i l}^{t} \right)}{(\vz_{\gilone}^{t})^2} \right)^{-1}  , \\
  \mx_{il}^{t} &= \vx_{il }^{t} \sum_{\gamma} \frac{F_{\gamma i}}{\vz_{\gilone}^{t}} \left( \mzp_{\gamma l \to i l}^{t} - \mz_{\gilone}^{t} \right) ,
\end{align}
and 
\begin{align}
 \mathbf{\mz}_{\mu l}^{t} &= \{ \mz_{\mu l}^t , \{ \mz_{\mu m}^{t-1} \}_{m \neq l} \} , \\
 \mathbf{\vz}_{\mu l}^{t} &= \{ \vz_{\mu l}^t , \{ \vz_{\mu m}^{t-1} \}_{m \neq l} \} .
\end{align}
Note that $\mxp_{il}$ is the MMSE estimator defined in (\ref{eq:MMSE}) and $\vxp_{il}$ is the variance of the local belief (\ref{eq:beliefsx}). 
$\mzp_{\mu l}$ and $\vzp_{\mu l}$ are defined in analogy.

We can then write the $\mzp_{\mu l \to il}$ as perturbations around $\mzp_{\mu l}$ using the relations~(\ref{eq:gderivatives}).
It is sufficient to compute the first order corrections with respect to the matrix elements $F_{\mu i}$, as those lead to corrections 
of order $1$ once summed. On the other hand, the corrective terms of higher order will remain of order $O(1/\sqrt{N})$ or smaller 
once summed, and do therefore not need to be explicitly calculated. This gives:
\begin{align}
 \mzp_{\mu l \to il}^t &= \gm( \mZ_{\muiltwo}^t, \vZ_{\muiltwo}^t, \Y_{\mu l} ) \nonumber  \\
		      &= \gm( \mZ_{\muiltwo}^t, \vZ_{\mu l}^t, \Y_{\mu l}) + O(F_{\mu i}^2) \nonumber \\
		      &= \mzp_{\mu l}^t + \mypartial{\mz_{\mu l}^t} \gm(\mZ_{\mu l}^t, \vZ_{\mu l}^t, \Y_{\mu l} ) \left( - F_{\mu i} \mxp_{il \to \mu l}^{t-1} \right) + O(F_{\mu i}^2)  \nonumber \\
		      &= \mzp_{\mu l}^t - F_{\mu i} \mxp_{il \to \mu l}^{t-1}  \frac{\vzp_{\mu l}^t}{\vz_{\mu l}^t} + O(F_{\mu i}^2) ,
\end{align}
and we can do the same for the  $\mxp_{il \to \mu l}$ messages, written as perturbations around $\mxp_{il}$ using the relations~(\ref{eq:fderivatives})
  \begin{align}
  {\mxp}_{il \to \mu l}^t &= \mxp_{il}^t + \mypartial{\mx} \fm^X({\mx}_{il}^t , {\vx}_{il}^t) \left( {\mx}_{il \to \mu l}^t - {\mx}_{il}^t \right) + O(F_{\mu i}^2) \nonumber \\
			  &=  {\mxp}_{il}^t + \frac{{\vxp}_{il}^t}{{\vx}_{il}^t} \left( - {\vx}_{il}^t \frac{F_{\mu i}\left( \mzp_{\mu l \to i l}^{t} - \mz_{\muilone}^{t} \right)}{\vz_{\muilone}^{t}}  \right) + O(F_{\mu i}^2) \nonumber \\
			  &=  {\mxp}_{il}^t - {\vxp}_{il}^t \frac{F_{\mu i}}{\vz_{\muilone}^{t}} \left( \mzp_{\mu l \to i l}^{t} - \mz_{\muilone}^{t} \right) + O(F_{\mu i}^2) .
  \end{align}
Using each of these equations in the other one, we obtain the perturbations:
\begin{align}
\mzp_{\mu l \to il}^{t} &= \mzp_{\mu l}^{t} - F_{\mu i}  \frac{\vzp_{\mu l}^t}{\vz_{\mu l}^t}  \mxp_{i l}^{t-1} + O(F_{\mu i}^2) , \\
 {\mxp}_{il \to \mu l}^t &= \mxp_{il}^t - F_{\mu i} \vxp_{il}^t \frac{\mzp_{\mu l}^t - \mz_{\mu l}^t}{\vz_{\mu l}^t} + O(F_{\mu i}^2) .
\end{align}
In the $N\to \infty$ limit, we therefore have
\begin{align}
 \vx_{il}^t &=  \left( \sum_{\gamma} \frac{F_{\gamma i}^2 ( \vz_{\gamma l}^t - \vzp_{\gamma l}^t)}{(\vz_{\gamma l}^t)^2} \right)^{-1}, \\
 \vz_{\mu l}^{t+1} &= \sum_{j} F_{\mu j}^2 \vxp_{j l}^{t}. 
\end{align}
This makes it possible to evaluate $\mz_{\mu l}$ and  $\mx_{i l}$ with only the local beliefs $\mxp_{il}$ and variances $\vxp_{il}$, such that 
in the $N \to \infty$ limit,
  \begin{align}
  {\mz}_{\mu l}^{t+1} &= \sum_i {F}_{\mu i } {\mxp}_{il}^t - \sum_i {F}_{\mu i}^2 {\vxp}_{il}^t \frac{\mzp_{\mu l}^t - \mz_{\mu l}^t}{\vz_{\mu l}^t}  ,\\
  {\mx}_{il}^{t+1} &= {\mxp}_{il}^t + {\vx}_{il}^{t+1} \sum_{\mu} {F}_{\mu i}   \frac{\mzp_{\mu l}^{t+1} - \mz_{\mu l}^{t+1}}{\vz_{\mu l}^{t+1}} .
  \end{align}
  With those steps made, we can greatly simplify the complexity of the
  message passing algorithm. The resulting version of algorithm \ref{algo:TAPMatrix} is 
  called ``TAP'' version, referring to the Thouless-Anderson-Palmer equations used in the study of spin glasses \cite{ThoulessAnderson77} with 
  the same technique.

%
%
%
    
\begin{algorithm}
\caption{Cal-AMP algorithm}
\label{algo:TAPMatrix}
  \textbf{Initialization:} for all indices $i$, $\mu$ and $l$, set
  \begin{align*}
  &{\mxp}_{il}^0 = 0, &{\mz}_{\mu l}^0 = \mzp_{\mu l}^0 = \y_{\mu l} , \\
  &{\vxp}_{il}^0 = \rho \sigma^2, &\vz_{\mu l}^0=\vzp_{\mu l}^0=1 \, .
  \end{align*}

  \textbf{Main loop:} while $t<t_{\rm max}$, calculate following quantities:
  \begin{align*}
    {\vZ}^{t+1} &= |\F|^2 {\vXp}^t \\
    {\mZ}^{t+1} &= {\F} {\mXp}^t - \vZ^{t+1} \odot \frac{\mZp^t - \mZ^t}{\vZ^t}  \\
    {\vZp}^{t+1} &= \gv \left( \vZ^{t+1}, \mZ^{t+1}, \Y \right) \\
    \mZp^{t+1} &= \gm \left( \vZ^{t+1}, \mZ^{t+1}, \Y \right) \\ 
   \vX^{t+1} &= \left(  \frac{(|\F|^2)^T \left( \vZ^{t+1} - \vZp^{t+1} \right)}{(\vZ^{t+1})^2} \right)^{-1} \\
   \mX^{t+1} &=  \mXp^{t} + \vX^{t+1} \odot \left(  \F^T \frac{ \mZp^{t+1} - \mZ^{t+1} }{\vZ^{t+1}} \right) \\
   {\vXp}^{t+1} &= \fm^X \left( {\mX}^{t+1} , {\vX}^{t+1}  \right) \\
   {\mXp}^{t+1} &=  \fv^X \left( {\mX}^{t+1} , {\vX}^{t+1}  \right) 
  \end{align*}
  \textbf{Result :} $\mxp_{il}^{t_{\rm max}}$ and ${\mz}_{\mu l}^{t_{\rm max}}$ are the estimates of $\x_{il}$ and $\z_{\mu l}$,
  and $\vxp_{il}^{t_{\rm max}}$ and ${\vz}_{\mu l}^{t_{\rm max}}$ are the uncertainties of those estimates.
\end{algorithm}

  Note that in this general version, we do not explicitly calculate
  estimates of $d_{\mu}$. The initialization can also be chosen using the probability distributions $\p_X$ and $\p_{Y|Z,D}$, but random 
  initialization provides good results.
  The use of the notations $\fm^X$, $\fv^X$, $\gm$ and $\gv$ is abusive and refers to their component-wise use in~(\ref{eq:beliefs}).  
  The algorithm remains valid for complex variables, in which case $(.)^T$ indicates complex transposition. 
  
  \subsection{Comparison to GAMP and perfectly calibrated GAMP}
  When $P=1$, Cal-AMP is strictly identical to GAMP, with:
  \begin{align}
   \gm(\mz,\vz,y) &= \frac{\mint{d_{\mu}} \p_D(d_{\mu}) f_1^Z(\mz,\vz,y,d)}{\mint{d_{\mu}} \p_D(d_{\mu}) f_0^Z(\mz,\vz,y,d)} , \\
   \gv(\mz,\vz,y) &= \frac{\mint{d_{\mu}} \p_D(d_{\mu}) f_2^Z(\mz,\vz,y,d)}{\mint{d_{\mu}} \p_D(d_{\mu}) f_0^Z(\mz,\vz,y,d)} - \gm(\mz,\vz,y)^2 . \nonumber
  \end{align}
  For $P>1$, the step involving $\gm$ and $\gv$ is the only one 
  in which the $P$ samples are not treated independently.
  
  If it is possible to perform perfect calibration of the sensors by supervised learning,
  one can replace the prior distribution $\p_D(d_{\mu})$ in the expressions for $\gm$ and $\gv$ 
  by $\delta(d_{\mu} - d_{\mu}^{\rm cal})$. In that case $\gm$ and $\gv$ can be calculated independently for the $P$ 
  samples, and Cal-AMP is once again identical to GAMP with perfectly calibrated sensors, which leads to:
  \begin{align}
   \gm(\mZ,\vZ,\Y) &= \gm(\mz_1,\vz_1,y_1) = \fm^Z(\mz_1, \vz_1, y_1, d^{\rm cal}) , \\
    \gm(\mZ,\vZ,\Y) &= \gv(\mz_1,\vz_1,y_1) = \fv^Z(\mz_1, \vz_1, y_1, d^{\rm cal}) .
  \end{align}
  Note that GAMP is usually written in a different way using
  \begin{align}
   g_{\rm out} = \frac{\mzp - \mz}{\vz} \quad \text{and} \quad   g_{\rm out}' &= \frac{\vzp - \vz}{\vz^2}.
  \end{align}

    \subsection{Damping scheme}
The stability of the algorithm can be improved with damping scheme proposed in~\cite{heskes2003stable}, 
which corresponds to damping the variances $\vz,\vx$ and the means $\mz,\mx$ with the following functions:
  \begin{align}
    {\rm var^{t+1}}  &\equiv (\beta \frac{1}{\rm var^{t+1}_0} + \frac{1-\beta}{\beta} \frac{1}{ \rm var^{t}})^{-1} ,\\
    \rm mean^{t+1}     &\equiv \beta' \rm mean^{t+1}_0 + (1-\beta') mean^t ,
  \end{align}
  where $\beta \in (0,1]$, $\beta'=\beta  {\rm var}^{t+1} / {\rm var_0}^{t+1}$ and the quantities with index $0$ are before damping. 
 

  \section{Examples of applications}
  In this section, we give two examples of how a sensor could introduce a distortion via the function $\p_{Y|Z,D}$.

  \subsection{Faulty sensors}
 
 In the non-CS case, the following setting has been studied before in the
  context of wireless sensor networks, for example in
  \cite{lo2013efficient,farruggia2011detecting}.
  For one signal sample $P=1$ this was also treated by GAMP in \cite{ziniel2014binary}.

  We assume that a fraction~$\epsilon$ of sensors is faulty (denoted by $d_{\mu}=0$) and only records noise $\sim \NN(\y_{\mu l};m_f,\sigma_f)$, 
  whereas the other sensors (with $d_{\mu}=1$) are functional and record $z_{\mu l}$.
   We then have
 \begin{align}
  \p_{Y|Z,D}(y|z,d) &= \delta(d-1) \delta(y-z) + \delta(d) \NN(y;m_f,\sigma_f) , \\
  \p_D(d) &= \epsilon \delta(d) + (1-\epsilon) \delta(d-1) ,
 \end{align}
 and this leads to analytical expressions for the functions $\gm$ and $\gv$, given in appendix~\ref{app:B}.

  If $m_f$ and $\sigma_f$ are sufficiently different from the mean and variance of the measurement taken by working sensors,
  the problem can be expected to be easy. But if $m_f$ and $\sigma_f$ are exactly the mean and variance of the measurements 
  taken by working sensors, nothing indicates which are the faulty sensors. The algorithm thus has to solve a problem of 
  combinatorial optimization consisting in finding which sensors are faulty. 

  \textbf{Perfect calibration:}
  If the sensors have been calibrated before, the problem can be 
  solved by a CS algorithm that discards the fraction $\epsilon$ of the measurements
  corresponding to the faulty sensors, 
leading to an effective measurement rate
  $\alpha_{\rm eff} = \alpha (1-\epsilon)$.
 The algorithm would then succeed in finding the solution if $\alpha_{\rm eff} > \alpha_{\rm CS}$. 
 Therefore a perfectly calibrated algorithm would have a phase transition at:
 \begin{equation}
  \alpha^{\rm cal}(\rho) = \alpha_{\rm CS}(\rho)/(1- \epsilon). \label{eq:alphaCalFaulty}
 \end{equation}

  Results of numerical experiments are presented on
  Fig.~\ref{faultySensors}, and show the comparisons with the perfectly calibrated case as well as the 
  increase in performance as the number of samples~$P$ grows.

  \subsection{Gain calibration}
  In this setting, studied in \cite{GribonvalChardon11,schulke2013blind}, each sensor multiplies the component $\z_{\mu l}$ by an unknown 
  gain $d_{\mu}^{-1}$. 
   One possible application is in the context of time-interleaved ADC converters, where gain calibration has been studied before \cite{saleem2011adaptive}.
  In noisy real gain calibration, the measurement process at each sensor is given by
  \begin{align}
  \y_{\mu l} = \frac{\z_{\mu l}+w_{\mu l}}{d_{\mu}},
  \end{align}
  with $w$ being Gaussian noise of mean $0$ and variance $\Delta$.
  Then the output channel is
  \begin{align}
  \p_{Y|Z,D}({\y}|{\z},{d}) &= \int {\rm d}w \g(w;0,\Delta) \delta(\y - \frac{\z - w}{d})  \nonumber \\
				    &= |d| \g(\z;d\y,\Delta) ,
  \end{align}
  and from this we can obtain that:
  \begin{align}
  f_0^Z(\mz_{\mu l},\vz_{\mu l},\y_{\mu l},d_{\mu}) 
    &\propto   |{d_{\mu}}| \g(d_{\mu};\frac{\mz_{\mu l}}{\y_{\mu l}},\frac{\Delta + \vz_{\mu l}}{|\y_{\mu l}|^2}).
  \end{align}
  
  This allows us to calculate $\gm$ and $\gv$, for which we obtain
\begin{align}
 \gm(\mZ_{\mu l}^t,\vZ_{\mu l}^t, y_{\mu m}) &= \frac{\Delta \vz_{\mu l}^t}{\Delta + \vz_{\mu l}^t} \left( \frac{\mz_{\mu l}^t}{\vz_{\mu l}^t} + \frac{y_{\mu l} \mdp_{\mu l}^t}{\Delta} \right) ,  \\
 \gv(\mZ_{\mu l}^t,\vZ_{\mu l}^t, y_{\mu m}) &= \frac{\Delta \vz_{\mu l}^t}{\Delta + \vz_{\mu l}^t} \left( 1 + \frac{\Delta \vz_{\mu l}^t}{\Delta + \vz_{\mu l}^t} \frac{y_{\mu l}^2}{\Delta^2} \vdp_{\mu l}^t \right) , \nonumber
\end{align}
  with
  \begin{align}
  \mdp_{\mu l}^t &= \fm^D(\md_{\mu l}^t,\vd_{\mu l}^t), \\
  \vdp_{\mu l}^t &= \fv^D(\md_{\mu l}^t,\vd_{\mu l}^t), \\
  \vd_{\mu l}^t  &=  \left( \sum_{m \neq l} \frac{|\y_{ \mu m}|^2}{\Delta + \vz_{\mu m}^{t-1}} + \frac{|\y_{ \mu l}|^2}{\Delta + \vz_{\mu l}^t} \right)^{-1} , \\
  \md_{\mu l}^t &= \vd_{\mu l}^t \left( \sum_{m \neq l} \frac{\mz_{\mu m}^{t-1} \y_{\mu m}^*}{\Delta + \vz_{\mu m}^{t-1}} + \frac{\mz_{\mu l}^{t} \y_{\mu l}^*}{\Delta + \vz_{\mu l}^{t}} \right),
  \end{align}
  where $\fm^D$ stands for $\fm^{|d|^P \p_D(d)}$.
   
  \textbf{Perfect calibration:}
  In this setting, if the sensors have been perfectly calibrated beforehand, the problem is equivalent to compressed 
  sensing, therefore 
  \begin{equation}
    \alpha^{\rm cal}(\rho) = \alpha_{\rm CS}(\rho). \label{eq:alphaCalGain}
  \end{equation}
  Another interesting lower bound for the necessary number of measures can be found. 
  Consider an oracle algorithm that
  knows the location of all the zeros in the signal, but not the calibration coefficients. 
  For each of the $M$ sensors, the $P$ measurements can be combined into $P-1$ independent equations of the type:
  \begin{equation}
   y_{\mu l} \sum_i F_{\mu i} x_{i m} - y_{\mu m} \sum_i F_{\mu i} x_{i l} = 0 
  \end{equation}
  There are $M(P-1)$ such linear equations and $P \rho N$ unknowns (as the algorithm knows all the zeros), therefore 
  it can find the solution only if $M(P-1)>P \rho N$, which leads to the lower bound:
  \begin{equation}
   \alpha_{\rm min}(\rho) = \frac{P}{P-1} \rho.
   \label{eq:alpha_min}
  \end{equation}

   \textbf{Complex gain calibration:}
   Cal-AMP also applies to the setting where $\mathbf{x}$, $\mathbf{F}$, $\mathbf{y}$ and $\mathbf{d}$ are complex
    instead of real. The algorithm is the same, with the difference that the update functions $f$ and $g$ calculated with 
    priors on complex numbers and with complex instead of real normal distributions.
 
  \section{Experimental results}
  
    Fig.~\ref{faultySensors} and~\ref{gainCalibration} show the results of numerical experiments made for the faulty sensors problem 
  and the gain calibration problem. All experiments were carried out on synthetic data and with priors matching the real signal distributions,
  \begin{equation}
   \p_X(\mathbf{\x}) = \prod_{il} \left[ (1-\rho) \delta(\x_{il}) + \rho \NN(\x_{il};0,1) \right] , \label{eq:bernouilligaussprior}
  \end{equation}
  and the corresponding update functions $\fm^X$ and $\fv^X$ have analytical
  expressions, given in appendix~\ref{app:B}.

  Effects of prior mismatch for CS has been studied in \cite{KrzakalaMezard12}, as well as the possibility to learn parameters of the priors with 
  expectation-maximization procedures. The measurement matrix was taken with random iid Gaussian elements with variance ${1}/{N}$, 
  such that $\mathbf{\z}$ is of order one,
  \begin{equation}
   \p_{\mathbf{F}}(\mathbf{F}) = \prod_{\mu i} \NN(F_{\mu i};0,\frac{1}{N}) .
  \end{equation}

  A MATLAB implementation of Cal-AMP algorithm~\ref{algo:TAPMatrix} was used. It is available at \url{github.com/cschuelke/CalAMP}. 
  For the priors used in the experiments, the integrals in $\fm$ and $\fv$ have simple 
  analytical expressions, and therefore the computational cost of the algorithm is dominated by matrix multiplications.

  In order to assess the quality of the reconstruction on synthetic data,
  we will look at the normalized cross-correlation between the generated and the reconstructed signal, $\mathbf{\x}$ and $\mathbf{\hat{x}}$: 
  used for instance in~\cite{gribonval2012blind,corbella2000analysis}:
   \begin{align}
  \mu (\mathbf{\x},\mathbf{\hat{x}}) &= \frac{1}{P} \sum_{l=1}^P \frac{ | \sum_{i=1}^N (x_{il} - \langle x_l \rangle )^* (\hat{x}_{il} - \langle \hat{x}_l \rangle) | }{ \sqrt{ \sum_{i=1}^N |x_{il} - \langle x_l \rangle |^2 \sum_{i=1}^N |\hat{x}_{il} - \langle \hat{x}_l \rangle |^2} },
  \end{align}
  where we have used the empirical means
  \begin{align}
   \langle x_l \rangle &= \frac{1}{N} \sum_i x_{il}  \quad \text{and} \quad \langle \hat{x}_l \rangle = \frac{1}{N} \sum_i \hat{x}_{il}.
  \end{align}
  Choosing this evaluation metric instead of the mean square error (MSE) allows to take into 
  account the fact that in some applications, there are ambiguities that are unliftable, in which case the MSE might be 
  a poor indicator of success and failure. 
  This is the case for complex gain calibration, where signal and calibration coefficients can only
  be recovered up to a global phase at best, and for real gain calibration in case of a mismatching prior $\p_D$.
  The normalized cross-correlation $\mu$ tends to $1$ for a perfect reconstruction,
  and it is therefore convenient to look at the quantity $\log_{10}(1-\mu)$.
  In all phase diagrams, the horizontal axis is the sparsity $\rho$ of the signal and the vertical axis is the measurement rate $\alpha$.

  \subsection{Faulty sensors}
  Fig.~\ref{faultySensors} shows the results of experiments made on the faulty sensors problem. 
  For a fraction $\epsilon$ of the sensors, the measurements are replaced by noise, such that if sensor $\mu$ is faulty, then
  \begin{equation}
   \p(\y_{\mu l}) = \NN(\y_{\mu l};m_f,\sigma_f),
  \end{equation}
  independently of $\z_{\mu l}$. In order to consider the hardest case, in which these measurements have the same distribution as $\z_{\mu l}$,
  we take the mean and variance to be
  \begin{align}
   m_f = 0  \quad {\rm and} \quad \sigma_f = \rho .
  \end{align}

  The results correspond well to the analysis
  made previously. GAMP can be applied and allows perfect reconstruction in some cases.
  However, using Cal-AMP and increasing $P$ allows to close the gap to the performances of a perfectly calibrated algorithm.
  
   \begin{figure}[h!]
  \centering
  \includegraphics[width=0.7\textwidth]{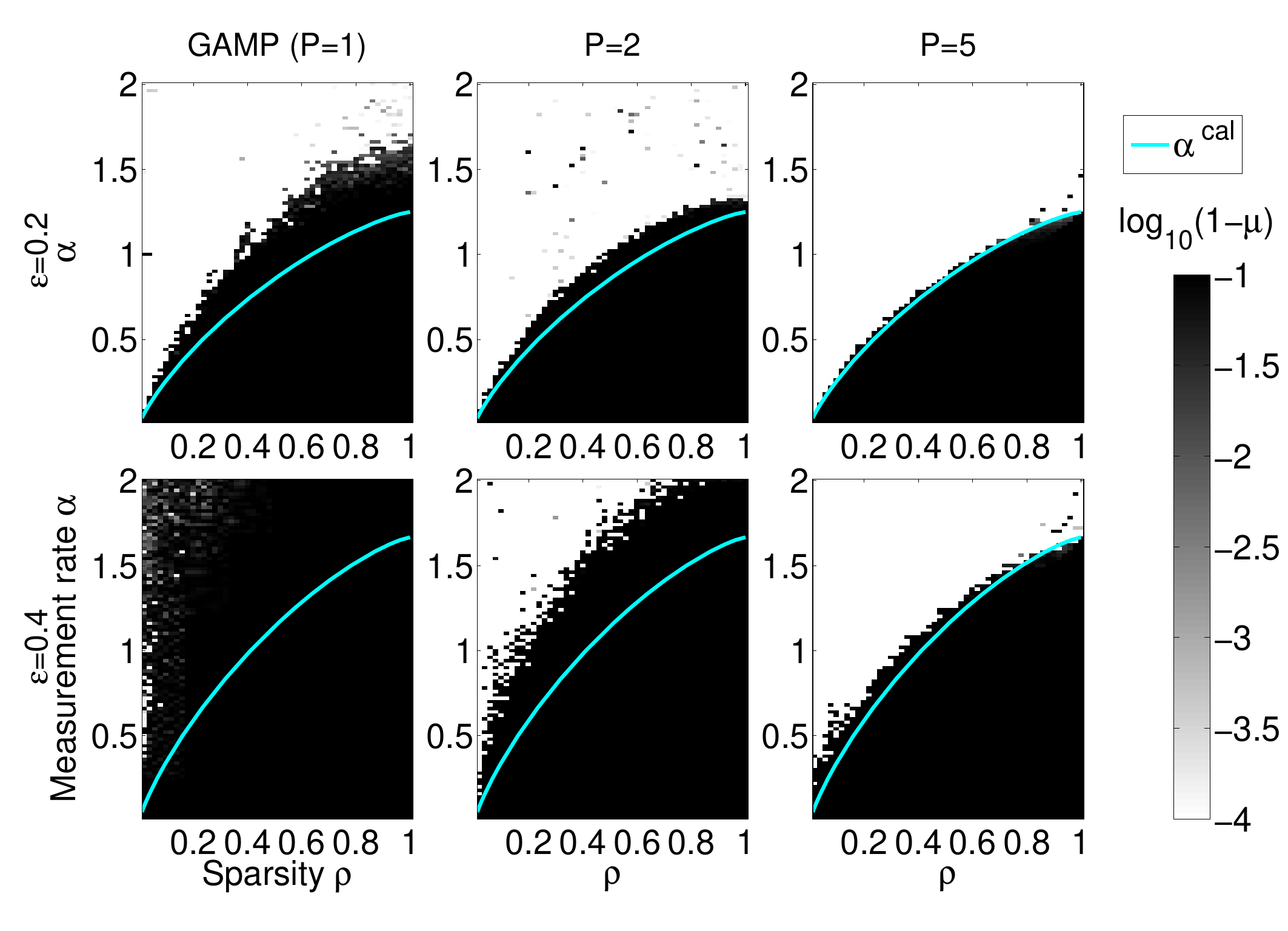}
  \caption{Phase diagrams for the faulty sensors problem. White indicates successful reconstruction, black indicates failure. Experiments were made for $N=1000$. In the upper row, the fraction of faulty sensors is $\epsilon=0.2$, while $\epsilon=0.4$ 
  in the lower row. 
  The line $\alpha^{\rm cal}$ from equation~(\ref{eq:alphaCalFaulty}) shows the performance of a perfectly calibrated algorithm.
  Increasing the number of samples $P$ allows to lower the phase
  transition down to $\alpha^{\rm cal}$, thus matching
  the performance of AMP algorithm knowing which sensors are faulty.}
  \label{faultySensors}
  \end{figure}


    \begin{figure*}
  \centering
  \includegraphics[width=0.9\textwidth]{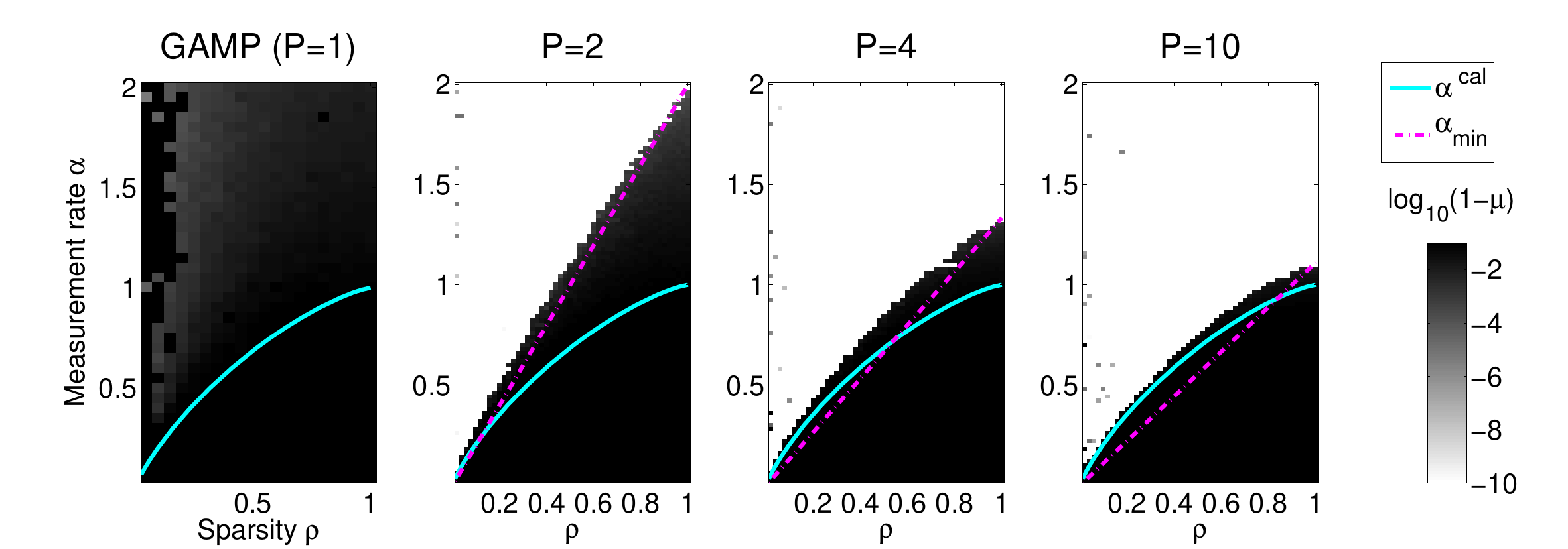}
  \caption{Phase diagrams for the real gain calibration problem.  White indicates successful reconstruction, black indicates failure.
  Experiments were made for $N=1000$ and $w_d=1$. 
  As the number of signals $P$ available for blind calibration increases, the lower bound $\alpha_{\rm min}(\rho)$
  from equation~(\ref{eq:alpha_min})  tends to $\rho$, 
  and the observed phase transition gets closer to $\alpha^{\rm cal}$, the transition of a perfectly calibrated algorithm.}
  \label{gainCalibration}
  \end{figure*}
  \subsection{Real gain calibration} 
  For the numerical experiments, the distribution chosen for the  calibration coefficients was a uniform distribution centered around $1$ and width $w_d<2$,
  \begin{equation}
   \p_D(d) = \UU(d;1,w_d).
   \label{eq:uniformprior}
  \end{equation}
    Experiments were made with a very low noise ($\Delta = 10^{-15}$), as taking $\Delta=0$ leads to occasional diverging behavior 
    of the algorithm. A damping coefficient $\beta=0.8$ was used, increasing the stability of the algorithm, while not slowing it 
    down significantly.
  \subsubsection{Bayes-optimal update functions}
  In that case, the update functions $f^D$ can be expressed
  analytically:
  \begin{align}
  \fm^D_{\UU} \left( R, \Sigma \right) &= \frac{I(P+1,R,\Sigma, \frac{2-w_d}{2}, \frac{2+w_d}{2})}{I(P,R,\Sigma, \frac{2-w_d}{2}, \frac{2+w_d}{2})} , \\
  \fv^D_{\UU} \left( R, \Sigma \right) &= \frac{I(P+2,R,\Sigma, \frac{2-w_d}{2}, \frac{2+w_d}{2})}{I(P,R,\Sigma, \frac{2-w_d}{2}, \frac{2+w_d}{2})} - \left( \fm^D_{\UU} \left( R, \Sigma \right) \right)^2, \nonumber
 \label{eq:bayesoptimalupdate}
 \end{align}
  with 
  \begin{align}
   I(&N,R,\Sigma,a,b) = \sum_{i=0}^N  \left[ \binom{N}{i} \frac{R^{N-i}}{2} (2 \Sigma)^{\frac{i+1}{2}}  \Gamma\left( \frac{i+1}{2} \right)  \right. \times \\
  &\left. \left( \sigma_{b}^i \gamma \left(\frac{i+1}{2}, \frac{(b-R)^2}{2 \Sigma} \right) -  \sigma_{a}^i \gamma \left(\frac{i+1}{2}, \frac{(a-R)^2}{2 \Sigma} \right)  \right) \right] , \nonumber
  \end{align}
 where $\Gamma$ is the gamma function, $\gamma$ is the incomplete gamma function
 \begin{align}
  \gamma(s,x) = \int_0^x t^{s-1} e^{-t} {\rm d}t,
 \end{align}
and $\sigma^i_x$ is $1$ if $i$ is even and the sign of $(x-R)$ if $i$ is uneven.

    Note that the fact that this prior has a bounded support can lead to a bad behavior of the algorithm.
  However, using a slightly bigger $w_d$ (by a factor $1.1$ in our implementation) in the prior
  than in the distribution used for generating $\mathbf{d}$ solves this issue.  

%
%

  \subsubsection{Results}
  Fig.~\ref{gainCalibration} shows the results in the case of the gain calibration problem. Here, signal recovery is impossible for $P=1$. 
  Furthermore, for $P>1$, the empirical phase transition closely matches the lower bounds 
  given by an uncalibrated oracle algorithm (\ref{eq:alpha_min}) and a perfectly calibrated algorithm (\ref{eq:alphaCalGain}). 
  Note that the exact position of the phase transition depends on the amplitude of the decalibration, given by $w_d$, as illustrated 
  on Fig.~\ref{gainCalibrationDetails}.
  
  Fig.~\ref{L1CalAMP} shows the comparison of performances of Cal-AMP with the algorithm
  relying on convex optimization used in \cite{GribonvalChardon11}. 
  Such an approach is possible in the case of gain calibration because the equation 
  \begin{equation}
   d_{\mu} \y_{\mu l} = \sum_i F_{\mu i} \x_{i l} + w_{\mu l}  \label{cvxeq}
  \end{equation}
  is convex both in $d_{\mu}$ and in $\x_{i l}$. 
  However, such a convex formulation is specific to this particular
  output model and is not generalizable to every type of sensor-induced distortion.
  The algorithm is implemented very easily using the CVX package \cite{cvx} by entering (\ref{cvxeq}) and adding an $L_1$
  regularizer on $\mathbf{x}$.
  The figure shows that Cal-AMP needs significantly less measurements for a successful reconstruction, and as shown in 
  Fig.~\ref{gainCalibrationDetails}, it is also substantially faster than its $L_1$ counterpart.

  \begin{figure*}[h!]
  \centering
  \includegraphics[width=0.4\textwidth]{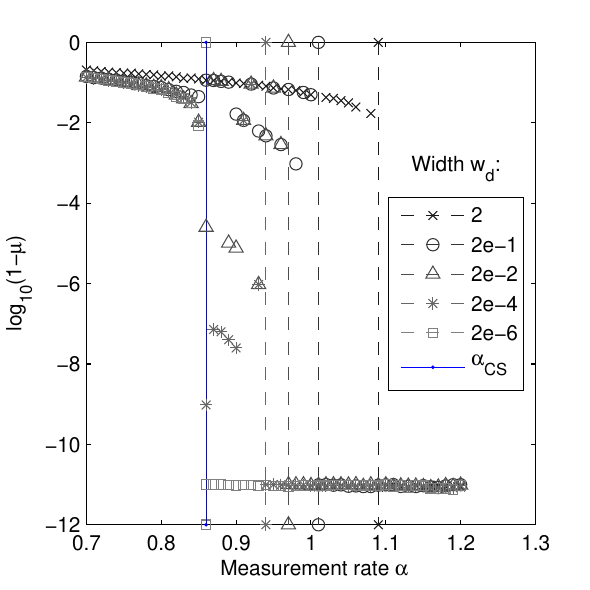}
  \includegraphics[width=0.4\textwidth]{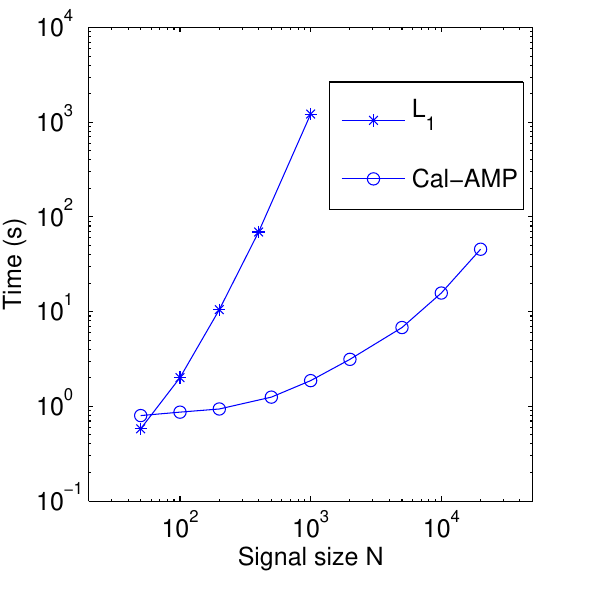}
  \caption{Left: Displacement of the phase transition  with varying decalibration amplitude $w_d$. Parameters are $\rho=0.7$, $P=4$ and $N=10000$. While for very small decalibrations, 
  the phase transition seems to be at the same location as in CS, it becomes clearly distinct with growing $w_d$.
  For each value of $w_d$, a vertical line materializes the empirical positions of the phase transitions (all points at the right of
  the line are perfectly reconstructed). 
  Right: Running times of Cal-AMP compared to the $L_1$ minimizing algorithm used in \cite{GribonvalChardon11}. 
  Both algorithms ran on a 2.4GHz processor, parameters were $\rho=0.2$, $\alpha=1$, $P=5$. Note that using structured operators,
  as Fourier transforms, can significantly reduce running times \cite{barbier2013compressed}. }
  \label{gainCalibrationDetails}
  \end{figure*}
  
  \begin{figure*}[h!]
  \centering
  \includegraphics[width=0.8\textwidth]{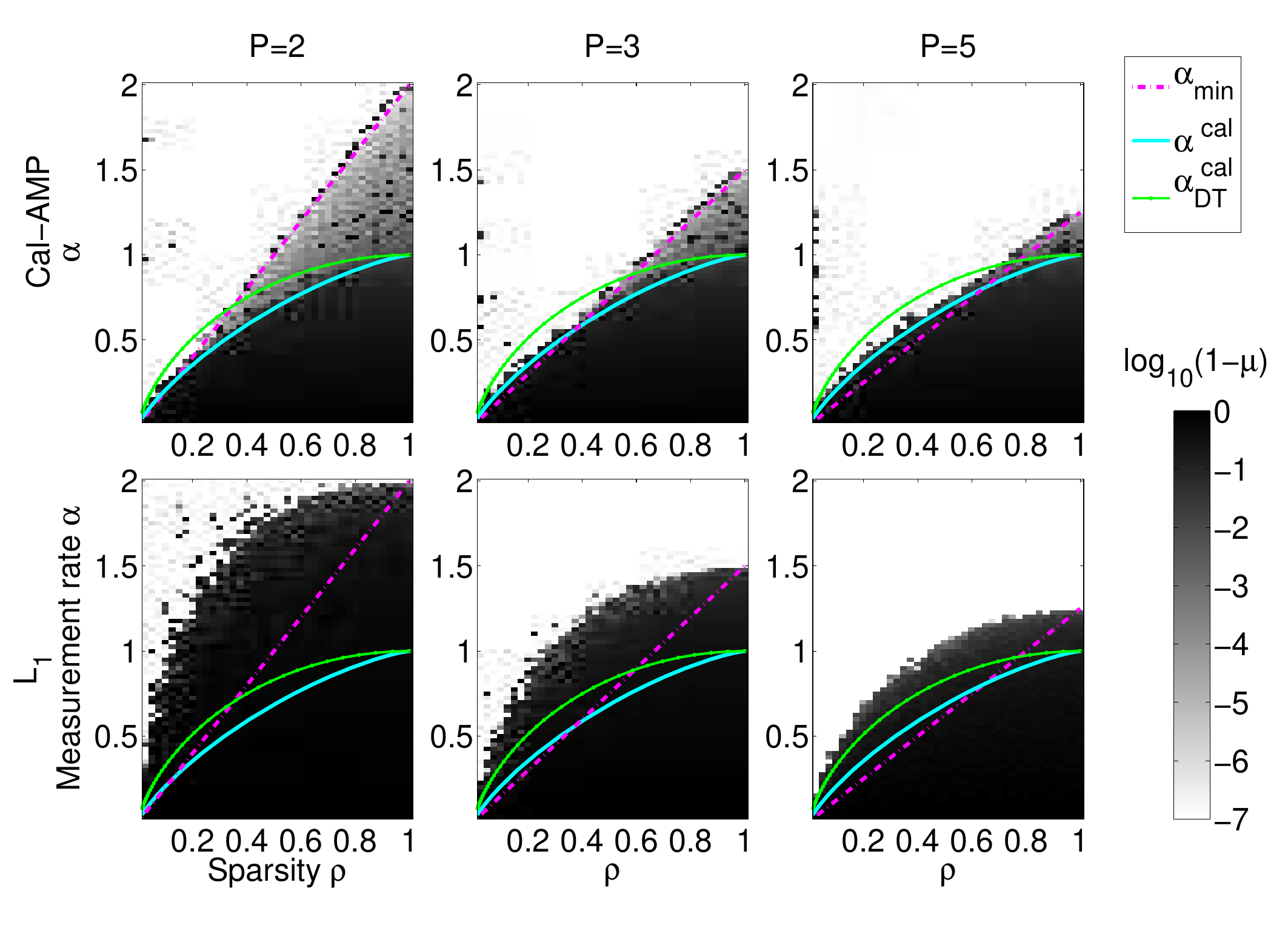}
  \caption{Experimental phase diagrams for Cal-AMP and an $L_1$-minimizing algorithm using the CVX package \cite{cvx}, for $N=100$ and $w_d=0.1$.
  While both algorithms show a similar qualitative behavior, Cal-AMP requires significantly less measurements for successful reconstruction (white region).
  The line $\alpha_{\rm min}$ is a lower bound from~(\ref{eq:alpha_min}), $\alpha^{\rm cal}$ is the phase transition of perfectly calibrated
  bayesian AMP, and  $\alpha^{\rm cal}_{\rm DT}$ is the Donoho-Tanner phase transition of a perfectly calibrated $L_1$-based
  CS algorithm \cite{Donoho05072005}. Just as the phase transition of Cal-AMP 
  approaches $\alpha^{\rm cal}$ with growing $P$, the one of the $L_1$ algorithm approaches $\alpha^{\rm cal}_{\rm DT}$. 
}
  \label{L1CalAMP}
  \end{figure*}

  \subsection{Complex gain calibration}
  For the numerical experiments, the distribution chosen for the calibration coefficients, the 
  signal and the measurement matrix use the complex normal distribution with mean $R$ and variance $\Sigma$,  which we note $\mathcal{CN}(x;R,\Sigma)$ :
  \begin{align}
   \p_X(x) &= (1-\rho)\delta(x) + \rho \mathcal{CN}(x;0,1), \label{eq:complexXdist} \\
   \p_F(F) &= \mathcal{CN}(F;0,1/N) , \label{eq:complexFdist} \\
   \p_D(d) &= \mathcal{CN}(d;0, 10 ) .   \label{eq:complexDdist}
  \end{align}
  The corresponding Bayes-optimal update functions $\fm^X$ and $\fv^X$ have analytical expressions~\cite{barbier2013compressed}, 
  given in appendix~\ref{app:B}.
  For the update functions $\fm^D$ and $\fv^D$, we use
  \begin{align}
   \fm^D(R,\Sigma) &= \frac{R}{|R|} \frac{I(P+1,|R|,\Sigma, 0, \infty)}{I(P,|R|,\Sigma, 0, \infty)} , \\
   \fv^D(R, \Sigma) &= \Sigma .
  \end{align} 
Though not Bayes-optimal, they lead to good results, presented in Figure~\ref{fig:ccal}.

\begin{figure*}
 \centering
 \includegraphics[width=0.9\textwidth]{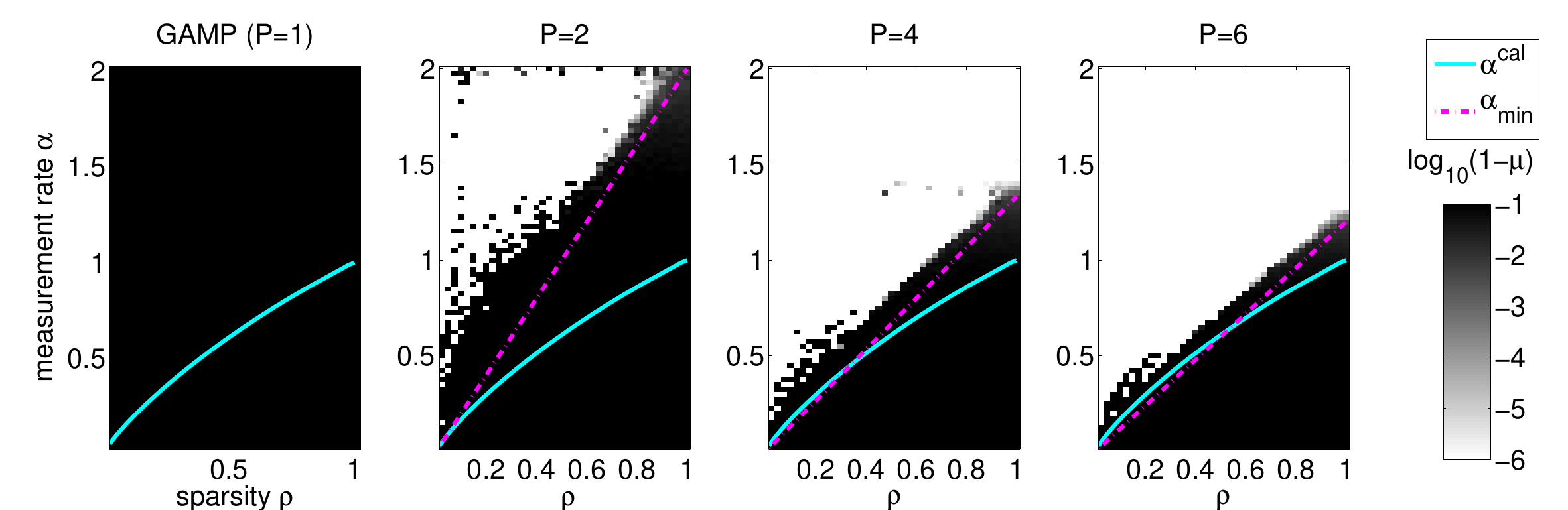}
 \caption{Experimental phase diagram for complex gain calibration using Cal-AMP for $N=500$. Here, $\alpha^{\rm cal}$ is the
 phase transition of a perfectly calibrated algorithm, that therefore performs as well as complex CS, analyzed in \cite{barbier2013compressed}.
 The line $\alpha_{\rm min}$ corresponds to the lower bound of (\ref{eq:alpha_min}).}
 \label{fig:ccal} 
\end{figure*}

  \section{Conclusion}
  In this paper, we have presented the Cal-AMP algorithm, designed for blind sensor calibration.
  Similar to GAMP, the framework allows to treat a variety of 
  different problems beyond the case of compressed sensing.
  The derivation of the algorithm was detailed, starting from the probabilistic formulation of the problem
  and the message-passing algorithm derived from belief propagation. 
  Two examples of problems falling into the Cal-AMP framework were studied numerically. Both for the 
  faulty sensors problem and the gain calibration problem, the performance of Cal-AMP was found to be close 
  to problem-specific lower bounds.

  Cal-AMP could find concrete applications in experimental setups using physical devices for data acquisition, in which 
  the ability to blindly calibrate the sensors might be either indispensable for good results, or allow 
  substantial cuts in hardware costs. 

  In compressed sensing the asymptotic behavior of the AMP algorithm
  was analyzed via the state evolution equations
  \cite{DonohoMaleki09,BayatiMontanari10}. We attempted to derive the
  corresponding theory for Cal-AMP, but even on the heuristic level
  the corresponding generalization turns out to be non-trivial. This
  analysis is hence left as an interesting open problem. 
 

 \appendix
 
 \section{Approximation of $\messx$}
 \label{app:A}
  We start by rewriting the messages~(\ref{eq:exactmesstildex}) using the function $g_0$ introduced in~(\ref{eq:gfunctions}):
  \begin{align}
   \tilde{\messx}_{\mu l \to il}^{t+1} (x_{il}) &\propto g_0( \mZ_{\muiltwo}^{t+1} + F_{\mu i} x_{il} \mathbf{e}_1, \vZ_{\muiltwo}^{t+1},\Y_{\mu}),  \label{eq:taylor_messx} 
  \end{align}
 where $\mZ_{\muiltwo}^{t+1}$ is a $P$-dimensional vector with first component $\mz_{\muilone}^{t+1}$, and its other components are
 $\mz_{\mu m}^t$ for $m \neq l$. The definition of $\vZ_{\muiltwo}^{t+1}$ is the same, replacing $\mz$ by $\vz$, and $\Y_{\mu l}$ is 
 the $P$-dimensional vector with first component $\y_{\mu l}$ and other components $\y_{\mu m}$ with $m \neq l$. Notice 
 that due to the definition of $g_0$, the order of the components $2$ to $P$ of those vectors is unimportant as long as it is the same 
 for each of them.
 $\mathbf{e}_1$ is the unit vector along the first direction of the $P$-dimensional space.
Making a Taylor expansion of~(\ref{eq:taylor_messx}), we obtain
  \begin{align}
    \tilde{\messx}_{\mu l \to il}^{t+1}(x_{il}) &\propto  {g_0( \mZ_{\mu l,i}^{t+1}, \vZ_{\mu l,i}^{t+1},\Y_{\mu})} \label{eq:expansion} \\
    &+  x_{il} F_{\mu i} \frac{\partial g_0( \mZ_{\mu l,i}^{t+1}, \vZ_{\mu l,i}^{t+1},\Y_{\mu})}{\partial \mz_{\mu l \to il}^{t+1}} \nonumber  \\
    &+ x_{il}^2 \frac{F_{\mu i}^2}{2}  \frac{\partial^2 g_0( \mZ_{\mu l,i}^{t+1}, \vZ_{\mu l,i}^{t+1},\Y_{\mu})}{\partial (\mz_{\mu l \to i l}^{t+1})^2}+ x_{il}^3 O(F_{\mu i}^3) . \nonumber
  \end{align}
 Let us now note that, for $a$ and $b$ of order one,
  \begin{align}
   \NN(x_{il};\frac{a}{ F_{\mu i} b},-\frac{1}{ F_{\mu i}^2 b}) &\propto  \NN(F_{\mu i} x_{il}; \frac{a}{b},-\frac{1}{b})  \label{eq:exp2}\\
               &\propto   e^{F_{\mu i} x_{il} a + ( F_{\mu i} x_{il})^2 \frac{b}{2} } \nonumber\\
               &\propto 1 + F_{\mu i} x_{il} a + \frac{(F_{\mu i} x_{il})^2}{2} ( b + a^2)  \nonumber \\
						      &+ O(F_{\mu i}^3 x_{il}^3).  \nonumber
  \end{align}
 We can now identify the coefficients of the expansion~(\ref{eq:expansion}) with those in~(\ref{eq:exp2}) to approximate
 the messages $\tilde{\messx}(x_{il}) $ as Gaussians around $F_{\mu i} \x_{il}=0$, with  mean $\mxa$ and variance $\vxa$:
\begin{equation}
  \tilde{\messx}_{\mu l \to il}^{t+1}(\x_{il}) \propto \NN(\x_{il}; \mxa_{\mu l \to il}, \vxa_{\mu l \to il}) + O(F_{\mu i}^3 x_{il}^3) ,
  \label{simple_tilde_m}
 \end{equation}
 were $\mxa$ and $\vxa$ have following expressions, found by expressing the derivatives of $g_0$ with the functions $\gm$ and $\gv$ 
 using the relations~(\ref{eq:gderivatives}):
 \begin{align}
  \vxa_{\mu l \to il} &= (\vz_{\muilone}^{t+1})^2 \left( F_{\mu i}^2 \left( \vz_{\muilone}^{t+1} - \vzp_{\mu l \to i l}^{t+1}\right) \right)^{-1} ,  \\
  \frac{\mxa_{\mu l \to il}}{\vxa_{\mu l \to il}} &= \frac{F_{\mu i}}{\vz_{\muilone}^{t+1}} \left( \mzp_{\mu l \to i l}^{t+1} - \mz_{\muilone}^{t+1} \right) .
 \end{align}
 This expression~(\ref{simple_tilde_m}) can now be used in~(\ref{mess_m}):
 \begin{align}
  \messx&_{il \to \mu l} (x_{il}) \propto \p_X(x_{il}) \prod_{\gamma \neq \mu} \tilde{\messx}_{\gamma l \to i l}(x_{il})  \label{eq:prodgauss} \\
			&\propto \p_X(x_{il}) \prod_{\gamma \neq \mu} \left( \NN(x_{il}; \mxa_{\gamma l \to il}, \vxa_{\gamma l \to il}) + O(F_{\gamma i}^3 x_{il}^3) \right) \nonumber \\
			&\propto \p_X(x_{il}) \prod_{\gamma \neq \mu} \NN(x_{il}; \mxa_{\gamma l \to il}, \vxa_{\gamma l \to il}) \prod_{\gamma \neq \mu} \left(1+ O(\frac{x_{il}^3}{N^{3/2}}) \right) \nonumber
 \end{align}
  The product of Gaussians that appears is proportional to another Gaussian. In fact,for any product of Gaussians,
  \begin{align}
  \prod_{k=1}^K \NN(x;R_k,\Sigma_k) &= \NN(x;R,\Sigma)  \frac{\prod_{k=1}^K \NN(R_k;0,\Sigma_k)}{\NN(R;0,\Sigma)} , 
  \label{prodOfGauss}
  \end{align}
  with 
  \begin{align}
  \Sigma^{-1} &= \sum_k \Sigma_k^{-1}  \, & \text{and} \,& &   R &= \Sigma \sum_k \frac{R_k}{\Sigma_k} .
  \label{newMeanAndVariance}
  \end{align}
 Moreover, the logarithm of the second product is $\sum_{\gamma \neq \mu} x_{il}^3 O(F_{\gamma i}^3) = x_{il}^3 O(1/\sqrt{N})$, 
 so the product is $1+x_{il}^3 O(1/\sqrt{N})$.
  The messages $\messx$ can therefore be written in the following way:
   \begin{equation}
  \messx_{\mu l \to il}^t ({\x}_{il}) \propto \p_X({\x}_{il}) \left( \NN({\x}_{il};\mx_{il \to \mu l}^t,\vx_{il \to \mu l}^t) + O(\frac{x_{il}^3}{\sqrt{N}}) \right), \nonumber
  \end{equation}
with
\begin{align}
  \vx_{il \to \mu l}^{t+1} &= \left( \sum_{\gamma \neq \mu} \frac{F_{\gamma i}^2 \left( \vz_{\gilone}^{t+1} - \vzp_{\gamma l \to i l}^{t+1} \right)}{(\vz_{\gilone}^{t+1})^2} \right)^{-1}, \nonumber \\
  \mx_{il \to \mu l}^{t+1} &= \vx_{il \to \mu l}^{t+1} \sum_{\gamma \neq \mu} \frac{F_{\gamma i}}{\vz_{\gilone}^{t+1}} \left( \mzp_{\gamma l \to i l}^{t+1} - \mz_{\gilone}^{t+1} \right). \nonumber
\end{align}

 \section{Analytical expressions of update functions}
 \label{app:B}
 \subsection{Faulty sensors problem} $\mzp$ and $\vzp$ are obtained from the functions $\gm$ and $\gv$ such that:
 \begin{align}
  \mzp_{\mu l} &= \frac{\epsilon \mz_{\mu l} \pi^f_{\mu} + (1-\epsilon) \y_{\mu l} \pi^z_{\mu}}{\epsilon \pi^f_{\mu} + (1-\epsilon) \pi^z_{\mu}} ,  \\
  \vzp_{\mu l} &= \frac{\epsilon (\mz_{\mu l}^2+\vz_{\mu l}) \pi^f_{\mu} + (1-\epsilon) |\y_{\mu l}|^2 \pi^z_{\mu}}{\epsilon \pi^f_{\mu} + (1-\epsilon) \pi^z_{\mu}} - |\mzp_{\mu l}|^2 , \nonumber 
 \end{align}
 with
 \begin{align}
 \pi^f_{\mu} &= \prod\limits_m \NN(y_{\mu m}; m_f, \sigma_f^2), \\
 \pi^z_{\mu} &= \prod\limits_m \NN(y_{\mu m}; \mz_{\mu m}, \vz_{\mu m}) .
 \end{align}

  \subsection{For Bernouilli-Gauss prior} 
  the update functions $\fm$ and $\fv$ corresponding to 
  the priors~(\ref{eq:bernouilligaussprior}) and~(\ref{eq:complexXdist}) can be found 
  in \cite{KrzakalaMezard12} and~\cite{barbier2013compressed} and obtained from:
   \begin{align}
  f_0^X(\mx,\vx) &= (1-\rho) \NN(\mx; 0,\vx) + \rho \NN(\mx; 0,\vx+1),  \nonumber \\
  f_1^X(\mx, \vx) &= \rho \frac{\mx}{\vx+1} \NN(\mx;0,\vx+1),  \\
  f_2^X(\mx,\vx) &= \rho \frac{ |\mx|^2 + \vx ( \vx+1)}{(\vx+1)^2} \NN(\mx; 0,\vx+1). \nonumber
  \end{align}
  In the complex case, all $\NN$ are replaced by $\mathcal{CN}$.





%

 
 \bibliographystyle{IEEEtran}
  \bibliography{refs.bib}

%








\end{document}